%Paper: hep-th/9404138
%From: Ali Chamseddine <chams@physik.unizh.ch>
%Date: Fri, 22 Apr 1994 11:12:52 +0200

%Definition for preprints
%
\magnification=\magstep1
\vsize=23truecm
\hsize=15.5truecm
\hoffset=.2truecm
\voffset=.8truecm
\parskip=.2truecm

\font\ti=cmbx10 scaled\magstep1
\font\eightrm=cmr8

\def\br{\hfill\break\noindent}

\def \ot {\otimes}
\def \op {\oplus}
\def \g5{\gamma_5}
\def \gm {\gamma^{\mu}}
\def \ra{\rightarrow}
\def \l4{\Bigl( {\rm Tr}( KK^*)^2-({\rm Tr}KK^*)^2\Bigr)}

\def \k2{{\rm Tr}KK^*}
\def \slash#1{/\kern -6pt#1}
\def \di{\slash{\partial}}

\pageno=0

%
% Titlepage
%
\baselineskip=.5truecm
\footline={\hfill}
{\hfill ETH/TH/94-12}
\vskip.2truecm
{\hfill 25 March 1994}
\vskip2.1truecm
\centerline{\ti Connection between Space-Time Supersymmetry }
\centerline{\ti and Non-Commutative Geometry}
\vskip1.2truecm
\centerline{  A. H. Chamseddine}
\vskip.8truecm
\centerline{ Theoretische Physik, ETH, CH-8093 Z\"urich Switzerland}
\vskip1.2truecm

\centerline{\bf Abstract}
\vskip.5truecm

\noindent
\vfill

\eject
\baselineskip=.6truecm
\footline={\hss\eightrm\folio\hss}

%section 1
%\centerline{}
\vskip1.1truecm
%{\bf\noindent 1. Introduction}
%\vskip.2truecm

\noindent
The non-commutative geometric construction of A. Connes [1-5]
has been successful in giving a geometrical interpretation
of the standard model as well as some grand unification models.
Our lack of ability to quantize the non-commutative actions
forced on us to quantize  the classical actions
resulting from the non-commutative ones by adopting  the usual
rules. The
symmetries that might be present in a non-commutative
action are then lost since they are not respected by the quantization
scheme. On the other hand, theories with space-time
supersymmetry has many desirable properties which are well
known [6]. It is then tempting to construct non-commutative
actions whose classical part has space-time supersymmetry.
Since all the particle physics models constructed using
non-commutative geometric methods correspond to
non-commutative spaces of a four-dimensional manifold
times a discrete set of points, it is natural to think
of extending this to a supermanifold times a discrete
set of points. However, it proved that there are many
mathematical issues that must be settled before
this approach could become acceptable.

An alternative way is to consider supersymmetric theories
in their component form. As supersymmetry transformations
relate the fermionic fields to the bosonic fields and vice
versa, it is possible to start with a fermionic action
to recover the bosonic one. The simplest example is
provided by the $N=1$ super Yang-Mills theory in
four dimensions . The
action is given by [7]:
$$
I =\int d^4x  \bigl(-{1\over 4} F_{\mu \nu}^aF^{\mu \nu a}
+{1\over 2} \overline{\lambda^a }\gm D_{\mu} \lambda^a \bigr),
\eqno(1)
$$
where $\lambda^a $ is a  Majorana spinor in the adjoint
representation of a gauge group G,
$F_{\mu \nu}^a $ is the field strength of the gauge field
$A_{\mu}^a$ and $D_{\mu }$ is a gauge covariant
derivative. The action (1) is invariant under the
supersymmetry transformations
$$\eqalignno{
\delta \lambda^a &=-{1\over 2} \gamma^{\mu \nu}F_{\mu \nu}^a
\epsilon ,& (2) \cr
\delta A_{\mu}^a &= \overline{\epsilon }\gamma_{\mu} \lambda^a ,
& (3) \cr }
$$
It is possible to derive the supersymmetric action (1) using
the Noether's method by
starting with the free fermionic part of (1) and the supersymmetry
transformations (2) and (3).

To reformulate the action (1) using the methods of
non-commutative geometry [1], we first define the triple
$({\cal A}, h, D )$ where h is the Hilbert space $L^2 (M, \tau,\sqrt
g d^4x)\ot C^n $
of spinors on a four-dimensional spin manifold $M$, ${\cal A}$
is the involutive algebra ${\cal A} =C^{\infty }(M) \ot M_n(C)$
of $n\times n$ matrix valued functions, and $D$ the Dirac operator
$D=\di \ot 1_n$ on $h$. The free part of the fermionic action
is written as
$$
{1\over 2}(\lambda , [\di ,\lambda ]),
$$
where $(, )$ denotes the scalar product on $L^2(S, \tau ,
\sqrt g d^4x )$ given by
$$
(\psi_1 ,\psi_2) =\int_M \sqrt g d^4x \tau \langle
\psi_1 (x), \psi_2 (x) \rangle , \eqno(4)
$$
where $\tau $ is a normalised trace on ${\cal A}$, and
$\langle , \rangle $ denotes the hermitian structure on
the left module ${\cal E}$ which in this paper will be
taken to be equal to ${\cal A}$. Let $\rho $ be a self-adjoint
element in the space $\Omega^1 ({\cal A})$ of one-forms:
$$
\rho =\sum_i a^i d b^i  \qquad (d1=0), \eqno(5)
$$
where $\Omega^.({\cal A})=\oplus \Omega^n ({\cal A}) $
is the universal algebra of differential forms. An involutive
representation of $\Omega^. ({\cal A})$ is provided by the map
$\pi :\Omega^. ({\cal A}) \ra B(h)$ defined by
$$
\pi (a_0 da_1 \ldots da_n) =a_0 [D,a_1][D,a_2]\ldots [D,a_n],
\eqno(6)
$$
where $B(h)$ is the algebra of bounded operators on $h$.
Then $\pi (\rho )=\sum a[D,b] $ is equal to $\gm A_{\mu} $
where $A_{\mu} =\sum a\partial_{\mu}b $. Since $\rho $ is
self-adjoint and $\gm $ is antihermitian, then $A_{\mu}^*
=-A_{\mu}$. The curvature of $\rho $ is $\theta =d\rho
+\rho^2 $ where $\theta \in \Omega^2 ({\cal A}) $. A
simple calculation shows that
$$
\pi (d\rho ) =\gamma^{\mu\nu}\partial_{\mu}A_{\nu}+
\sum \partial^{\mu}a \partial_{\mu}b ,\eqno(7)
$$
If $\pi (\rho ) \in {\rm Ker (\pi )}$, then
$$\pi (d\rho )
=\sum \partial^{\mu }a\partial_{\mu}b =-\sum a \partial^{\mu}\partial_
{\mu} b , \eqno(8)
$$
is an independent scalar function. The choice of $\pi (d\rho )$ in
$\pi (\Omega^2 ({\cal A}))\setminus \pi (d {\rm Ker} \pi
\setminus_{\Omega^1({\cal A})})$ is uniquely determined to be
orthogonal to all auxiliary fields, with respect to the inner
product on $\Omega^2 ({\cal A})$. From this we deduce that,
modulo the auxiliary field (i.e. the kernel of $\pi (d\rho )$ ),
$\pi (\theta ) = \gamma^{\mu\nu} F_{\mu\nu}$. The
Yang-Mills action is
$$\eqalign{
{1\over 2}{\rm Tr}_w (\theta ^2 D^{-4})&=
{1\over 2}\int \sqrt g d^4x {\rm Tr} (
\pi (\theta )^2 ) \cr
&=\int \sqrt g d^4x {\rm Tr} (-{1\over 4} F_{\mu\nu}F^{\mu\nu}),
\cr }\eqno(9)
$$
where ${\rm Tr}_w $ is the Dixmier trace [1].
The interacting fermionic action is
$$
{1\over 2}(\lambda ,[D+\rho ,\lambda]) ={1\over 2}
\int \sqrt g d^4x {\rm Tr}
(\overline{\lambda} \gm [\partial_{\mu} +A_{\mu}, \lambda ]). \eqno(10)
$$
After analytically continuing from Euclidean to Minkowski space
by the change $x_4\ra it $,
the action given by the sum of (9) and (10) changes by $I_E\ra -I_M$.
 The supersymmetry transformation for
$\lambda $ takes the simple form
$$
\delta \lambda =-\pi (\theta )\epsilon  , \eqno(11)
$$
while for $\rho $ it is given by
$$
\delta \pi (\rho )=\overline{\epsilon } E_a \lambda (E_a ),\eqno(12)
$$
where $E_a $ is a local orthonormal basis of $\Omega_D^1 ({\cal A})
\equiv \Omega^1 ({\cal A})\setminus ({\rm Ker}\pi +d {\rm Ker}\pi )$.
In our case the basis is $E^a =\gamma^a $.

Since we still do not know how
to construct in the non-commutative framework  the most general
 supersymmetric action, we shall only address the problem
of finding  supersymmetric
theories which also correspond to  non-commutative spaces, in the
same sense that the standard model has such an interpretation.
In reality
we will show that not all supersymmetric theories do correspond
to non-commutative spaces.

We next consider the $N=2$ super Yang-Mills action [8]. It is
given by
$$\eqalign{
I&=\int d^4x  \Bigl( -{1\over 4} F_{\mu\nu}^aF^{\mu\nu a}
+{1\over 2} D_{\mu}S^aD^{\mu}S^a +{1\over 2} D_{\mu}P^aD^{\mu}P^a
+\overline{\chi }^a\gm D_{\mu} \chi^a  \cr
&\qquad -if^{abc}\overline{\chi }^a (S^b-i\g5 P^b)\chi^c -{1\over 2}
 \bigl( f^{abc}S^bP^c  \bigr)^2 \Bigr), \cr }\eqno(13)
$$
where  $S^a$ and $P^a$ are a scalar and pseudoscalar fields,
and $\chi^a $
is a Dirac spinor, all in the adjoint representation of the
gauge group. The action (13) is invariant under the
transformations:
$$\eqalign{
\delta A_{\mu}^a &=\overline{\epsilon} \gamma_{\mu}\chi^a
-\overline{\chi^a} \gamma_{\mu}\epsilon ,\cr
\delta P^a&=\overline{\chi^a }\g5 \epsilon -\overline{\epsilon}
\g5 \chi^a ,\cr
\delta S^a&= i(\overline{\chi^a}\epsilon -\overline{\epsilon}\chi^a ),\cr
\delta \chi^a &=\bigl(-{1\over 2} \gamma^{\mu\nu}F_{\mu\nu}^a
-\g5 f^{abc}P^bS^c +i\gm (D_{\mu}S^a-i\g5
D_{\mu}P^a)\bigr)\epsilon . \cr }
\eqno(14)
$$
{}From our experience with the non-commutative construction
of the standard model, and since in the action (13) a
complex scalar field is unified with a gauge field, an
obvious guess is to take the non-commutative space to be
$M_4\times ({\rm two\ points})$, with the algebra
$$
{\cal A}=C^{\infty}(M_4)\otimes M_n(C) \oplus C^{\infty}(M_4)
\otimes M_n(C),  \eqno(15)
$$
and the Dirac operator
$$
D=\pmatrix{
\di \ot 1_n &i\g5 \ot\phi_0 \cr
-i\g5 \ot \phi_0^* & \di \ot 1_n \cr },\eqno(16)
$$
acting on the Hilbert space of spinors of the form
$$
\lambda =\pmatrix{ L\chi \cr R\chi \cr}, \eqno(17)
$$
where $L={1\over 2} (1+\g5)$ and $R={1\over 2} (1-\g5)$, and
$\chi $ is a Dirac spinor. Elements of ${\cal A}$ are taken
to be operators of the form $\pmatrix{a&0\cr 0&a}$ where $a $ is
a smooth function on $M_4$ with values in $M_n(C)$. The parameters
$\phi_0 $ appearing in eq. (16) are taken to be arbitrary except
for the constraint
$$ [\phi_0 ,\phi_0^* ] =0.  \eqno(18)
$$
A self-adjoint element $\rho $ in the space $\Omega^1({\cal A}) $
has the representation
$$
\pi (\rho )=\pmatrix{ \gm A_{\mu} &i\g5 \phi \cr -i\g5 \phi^* & \gm
A_{\mu} \cr} ,\eqno(19)
$$
where $A_{\mu}=\sum a\partial_{\mu}b $, $\phi +\phi_0 =\sum a
\phi_0 b $ and $\phi^* +\phi_0^* =\sum a \phi_0^* b $. We have
assumed, without any loss in generality, that $\sum ab =1$. The
fermionic action in (13) can  now be simply written as
$$
{1\over 2}\bigl(\lambda ,[D+\pi (\rho ) ,\lambda ]\bigr)=
{1\over 2}\int \sqrt g d^4x
{\rm Tr} \bigl( \overline{\lambda} [D+\pi (\rho ),\lambda ]
\bigr). \eqno(20)
$$
We must now prove that the curvature square of $\rho $
constructed with the Dirac operator (16) yields the
correct bosonic part of the action (13). First we compute
$\pi (d\rho )=\sum [D,a][D,b] $ which can be represented as
a $2\times 2$ matrix whose elements are
$$\pi (d\rho )=\pmatrix{ \gamma^{\mu\nu}\partial_{\mu} A_{\nu} +X
& i\gm \g5 (\partial_{\mu}\phi +[A_{\mu},\phi_0 ]  \cr
-i\gm \g5 (\partial_{\mu}\phi^* +[A_{\mu},\phi_0^*] &
\gamma^{\mu\nu}\partial_{\mu}A_{\nu} +X' \cr},\eqno(21)
$$
where
$$\eqalign{
X&=\sum \partial^{\mu }a\partial_{\mu}b +\phi_0\phi^*
+\phi\phi_0^* -\sum a[\phi_0\phi_0* ,b] \cr
X'&=\sum \partial^{\mu}a\partial_{\mu}b +\phi_0^*\phi +\phi^*
\phi_0 -\sum a [\phi_0^*\phi_0 ,b] \cr}.\eqno(22)
$$
If $\rho \in {\rm Ker}\pi $ then $\pi (d\rho )$ will be given
$$\eqalign{
\pi (d\rho )\vert_{\pi (\rho )=0}&=\sum \bigl(  \partial^{\mu }a
\partial_{\mu}b - a[\phi_0\phi_0^* ,b] \bigr) \ot 1_2 ,\cr} \eqno(22)
$$
where we have used the constraint eq. (18). This does not
constitute any loss of generality since $\phi_0 $ will decouple
from the final action. The curvature  $\theta =d\rho +\rho^2 $,
after moding out by the kernel of $\pi (d\rho )$, is:
$$\eqalign{
\pi (\theta )_{11} &={1\over 2}\gamma^{\mu\nu}F_{\mu\nu}
+{1\over 2}[(\phi +\phi_0), (\phi^* +\phi_0^*)] \cr
\pi (\theta )_{12}&=i\gm\g5 \bigl( \partial_{\mu}(\phi +\phi_0 )
+[A_{\mu} ,(\phi +\phi_0 )]\bigr) \cr
\pi (\theta )_{12}&=-i\gm\g5 \bigl( \partial_{\mu}(\phi^* +\phi_0^*)
+[A_{\mu}, (\phi^* +\phi_0^*)] \bigr) \cr
\pi (\theta )_{22} &={1\over 2}\gamma^{\mu\nu}F_{\mu\nu}
 -{1\over 2} [(\phi +\phi_0),
(\phi^* +\phi_0^*]. \cr }\eqno(23)
$$
Notice that the potential terms in $\pi (\theta )_{11}$ and
$\pi (\theta )_{22}$ have opposite signs as the diagonal part of
$ \pi (\theta )$ has been moded out. By redefining $\phi +\phi_0
\ra  \phi $ one sees that $\phi_0 $ drops completely from eq. (23).
   The bosonic part of the
non-commutative action is then given by
$$\eqalign{
{1\over 4} {\rm Tr}_w \bigl( \theta^2 D^{-4}\bigr) &=\int \sqrt
g d^4 x {\rm Tr} \bigl( -{1\over 4} F_{\mu\nu}F^{\mu\nu}
-{1\over 2}D_{\mu}\phi D^{\mu}\phi
 +{1\over 8} ([\phi ,\phi^*])^2 \bigr). \cr }\eqno(24)
$$
Continuing from Euclidean to Minkowski space and inserting
$\phi =S-iP $, we exactly recover the bosonic part of the
supersymmetric action (13). The supersymmetry transformations
are now very simple:
$$\eqalign{
\delta \lambda &=-\pi (\theta )\epsilon \cr
\delta \pi (\rho )&= \bigl( \overline{\epsilon }E_i \lambda
-\overline{\lambda }E_i \epsilon \bigr) E_i ,\cr}\eqno(25)
$$
where $E_i$ a local orthonormal basis of $\Omega_1 ({\cal A})$,
and $\epsilon $ has the same representation as $\lambda $
in eq. (17).
In this case the basis can be taken to be
$$\eqalign{
E_a&= \gamma_a \ot 1_2 \cr
E_5&= i\g5 \ot \tau_1 \cr
E_6&= i\g5 \ot \tau_2 ,\cr} \eqno(26)
$$
where $\tau_1$ and $\tau_2 $ are Pauli matrices. It is easy
to see that the transformations in (14) agree completely  with
those in (25) by substituting $\chi =L\chi +R\chi $ and
similarly for $\epsilon $. The dimension of the module in
this example is six, since this is the number of independent
elements in the basis.

Another non-trivial example is provided by the $N=4$ super
Yang-Mills action [9]. This is given by
$$\eqalign{
I&=\int \sqrt g d^4 x  \bigl( -{1\over 4}F_{\mu\nu}^a
F^{\mu\nu a} +{1\over 2}D_{\mu}\phi_{ij}^aD^{\mu}\phi^{ija}
 +\overline{\chi^{ia}}\gm D_{\mu}L\chi^{ia} \cr
&\qquad  -{i\over 2} \overline{\tilde {\chi^{ia}}}\phi_{ij}^a, L\chi^{ja}
+{i\over 2}f^{abc} \overline{\chi^{ia}} \phi^{ijb} R\chi_j^c
 -{1\over 4}(f^{abc}\phi_{ij}^b\phi_{kl}^c )(f^{ade}\phi^{ijd},\phi^{kle})
\bigr), \cr }\eqno(27)
$$
where $\tilde{\chi_i} =C\overline{\chi^i}^T $ is the conjugate
spinor, $i=1,\cdots ,4$, $\chi^i $ transforms as a $4$ and
$\chi_i $ transforms as a $\overline{4}$ of SU(4). The scalars
$\phi_{ij} $ are self-dual ,$\phi_{ij}={1\over 2}
\epsilon_{ijkl} \phi^{kl} $, transforming as a $6$ of SU(4).
The action (27) is invariant under the supersymmetry transformations
$$\eqalign{
\delta A_{\mu}^a&=\overline{\epsilon}_i\gamma_{\mu}L\chi^{ia}
-\overline{\chi}_i^a \gamma_{\mu}L\epsilon^i \cr
\delta \phi_{ij}^a&= i(\overline{\epsilon}_j R\tilde{\chi}_i^a
-\overline{\epsilon}_i R \tilde{\chi}_j^a +\epsilon_{ijkl}
\overline{\epsilon}^k L\chi^{la} )\cr
\delta (L\chi^{ia}) &=-{1\over 2} \gamma^{\mu\nu}F_{\mu\nu}^a
L\tilde{\epsilon}^i +i\gm D_{\mu}\phi^{ija}R\tilde{\epsilon}_j
+{1\over 2}f^{abc}\phi^{ikb} \phi_{kj}^c L\epsilon^j \cr
\delta (R\tilde{\chi}_i^a) &=-{1\over 2}\gamma^{\mu\nu}F_{\mu\nu}^a
R\tilde{\epsilon}_i -i\gm D_{\mu}\phi_{ij}^aL\epsilon^j
+{1\over 2} f^{abc}\phi_{ik}^b\phi^{kjc}R\tilde{\epsilon}_j .\cr}\eqno(28)
$$
In order to rewrite the free fermionic interactions of (27) in the
form ${1\over 2}\overline{\lambda } [D,\lambda ]$ for a generalized Dirac
operator $D$ we define
$$
\lambda =\pmatrix{L\chi^i \cr R \tilde{\chi}_i\cr},
$$
and we take the algebra ${\cal A}$  to be ${\cal A }=C^{\infty}(M_4)
\ot M_n(C) $. The Dirac operator is
$$
D=\pmatrix{\di \ot 1_4 & i\g5 \ot \phi_{0ij} \cr
-i\g5\ot \phi_0^{ij} & \di \ot 1_4 \cr},\eqno(29)
$$
where we have taken $\phi_{0ij}$ to be self-dual constant matrices.
The Hilbert space is $L^2 \ot \sum_{l=1}^8  M_n^l(C)$, and the
involutive representation $\pi (a) $ is given by
$$\eqalign{
\pi (a) &=\bigl( \pi_0 (a)\op \pi_0 (a)\op \pi_0 (a) \op \pi_0
(a)\bigr) \cr &\qquad \op \bigl( \overline{\pi}_0 (a) \op
\overline{\pi}_0 (a) \op \overline{\pi}_0 (a) \op \overline{\pi}_0
(a) \bigr), \cr}\eqno(30)
$$
so that $a\in {\cal A}$ has the representation
$$
a\ra {\rm diag} (a,a,a,a,\overline{a},\overline{a},\overline{a},
\overline{a} ), \eqno(31)
$$
where $\overline{a} $ denotes the complex conjugate of $a$. A one-form
 $\rho =\sum a db $ in $\Omega^1 ({\cal A}) $ has the representation
$$
\pi (\rho )=\pmatrix{\gm A_{\mu} \ot 1_4 &i\g5 \ot \phi_{ij} \cr
-i\g5 \ot \phi^{ij} & \gm \overline{A}_{\mu} \ot 1_4 \cr},\eqno(32)
$$
where $\phi_{ij}+\phi_{0ij}=\sum a \phi_{0ij}\overline {b} $ is
self dual, and $\phi^{ij}+\phi_0^{ij}=\sum \overline{a}
\phi_0^{ij} b$. In analogy with the previous case, and after
moding out by the kernel of $\pi (d\rho )$ which is diagonal,
one finds that the curvature matrix has the components
$$\eqalign{
\pi (\theta )_{11}&={1\over 2}\gamma^{\mu\nu}F_{\mu\nu}\ot
\delta_i^j +{1\over 2} [(\phi +\phi_0),(\overline {\phi}
+\overline{\phi }_0)]_i^j \cr
\pi (\theta )_{12}&= i\gm\g5 \bigl( \partial_{\mu}(\phi +\phi_0 )_{ij}
+[A_{\mu}, (\phi +\phi_0 )]_{ij} \bigr) \cr
\pi (\theta )_{21}&= -i\gm\g5 \bigl( \partial_{\mu}(\phi +\phi_0)^{ij}
+[A_{\mu}, (\phi +\phi_0 )^{ij}\bigr) \cr
\pi (\theta )_{22}&= {1\over 2}\gamma^{\mu\nu}F_{\mu\nu}\ot \delta_i^j
-{1\over 2}[(\phi +\phi_0 ), (\overline{\phi} +\overline{\phi}_0)]_i^j.
\cr}\eqno(33)
$$
After redefining $\phi +\phi_0 \ra \phi $ the
bosonic action becomes:
$$\eqalign{
I_b&={1\over 4}{\rm Tr}\bigl( \theta^2 D^{-4}\bigr) \cr
  &=\int \sqrt g d^4 x {\rm Tr} \bigl( -{1\over 4}F_{\mu\nu}F^{\mu\nu}
-{1\over 2} D_{\mu}\phi_{ij}D^{\mu}\phi^{ij} +{1\over 8}
([\phi ,\overline{\phi }][\phi ,\overline{\phi }])_i^i \bigr). \cr}
\eqno(34)
$$
After analytically continuing to Minkowski space and using
the identity
$${\rm Tr}([\phi ,\overline{\phi}][\phi ,\overline{\phi}])
={\rm Tr}([\phi, \phi ][\overline{\phi },\overline{\phi}]),
$$
valid for a self-dual field $\phi $, we find that
the action  (34) agrees completely with the bosonic part of
(27). The fermionic action is expressed, as before, in the form
$$
{1\over 2}\bigl( \overline{\lambda }, [D+\pi (\rho ),\lambda ]\bigr),
\eqno(35)
$$
and this reproduces the fermionic part of (27).
The supersymmetry transformations are also simplified
to the form
$$\eqalign{
\delta \lambda &= -\pi (\theta )\epsilon \cr
\delta \pi (\rho )&= \bigl( \overline{\epsilon} E_A \lambda
 \bigr) E_A, \cr }\eqno(36)
$$
where $\epsilon $ has the same representation as $\lambda $ and
$E_A$ is an orthonormal basis of $\Omega_D^1 ({\cal A})$. In this
case it is given by
$$\eqalign{
E_a&= \gamma_a \ot 1_8 \cr
E_{ij}&= i\g5 \ot \pmatrix{0& \rho^{ij}\cr \rho_{ij} &0 \cr},\cr}\eqno(37)
$$
where $(\rho^{ij})_{kl}=\delta_k^i \delta_l^j -\delta_l^i
\delta_j^k $ and $(\rho_{ij})_{kl}=\epsilon_{ijkl} ={1\over 2}
\epsilon_{ijmn} (\rho^{mn})_{kl} $. The dimension of the module
is ten. It is a well known fact that $N=2$ and $N=4$ super Yang-Mills
theory in four dimensions can be obtained by dimensional reduction
of $N=1$ super Yang-Mlls theory in six and ten dimensions
respectively [9].
For these higher dimensional theories,
the fermionic action is of the form $\overline{\lambda} [\Gamma^M
D_M ,\lambda ]$ where $\Gamma^M $ are Dirac matrices in the
respective dimension. It is remarkable that one can interpret
the four dimensional theories as corresponding to non-commutative
spaces, and where the non-commutative construction yields the
same answers as the known higher dimensional theories.

It is important
to determine which of the $N=1$ supersymmetric theories correspond to
actions of non-commutative spaces. A general globally
$N=1$ super Yang-Mills theory coupled to $N=1$ supersymmetric
matter is [7,6]:
$$\eqalign{
I&=\int d^4 x \sqrt g \Bigl(
 -{1\over 4}F_{\mu\nu}^aF^{\mu\nu a} - {1\over 2}\overline{\lambda}^a
\gm D_{\mu} \lambda^a  +D_{\mu}z_i^* D^{\mu }z^i \cr &\qquad
- \bigl(\overline{\chi}^i \gm D_{\mu}\chi^i -2i\overline{\chi_L^i}
 \lambda^a (T^az)^i +{\rm h.c}\bigr) -{1\over 2}
\bigl(z_i ^* (T^az)^i +\xi^a \bigr)^2 \Bigr), \cr}\eqno(38)
$$
where
$$\eqalign{
D_{\mu}z^i =\partial_{\mu}z^i +iA_{\mu}^a (T^a)_i^j z^j \cr
D_{\mu}\chi^i =\partial_{\mu}\chi^i +iA_{\mu}^a (T^a)_i^j \chi^j
\cr
D_{\mu}\lambda^a =\partial_{\mu}\lambda^a -g f^{abc}A_{\mu}^b
\lambda^c \cr
[T^a ,T^b]=if^{abc} T^c \cr
{\rm Tr}(T^a T^b)=\tau_R \delta^{ab} \cr
}\eqno(39)
$$
and $T^a $ are the generators of the gauge group G with structure
constants $f^{abc}$ and $\xi^a $ are constants associated with
the abelian generators of G.
The chiral multiplet $(z^i ,\chi_L^i )$ is in some representation,
usually reducible, of the gauge group.
 This action is invariant under the
supersymmetry transformations:
$$\eqalign{
\delta A_{\mu}^a &=\overline{\epsilon}\gamma_{\mu}\lambda^a \cr
\delta \lambda^a &=-{1\over 2} \gamma^{\mu\nu}F_{\mu\nu}\epsilon
+i(z^*T^a z +\xi^a )\g5\epsilon \cr
\delta z^i &=2 \overline{\epsilon} \chi_L^i \cr
\delta \chi_L^i &= (\gm \epsilon_R )D_{\mu} z^i .\cr }
\eqno(40)
$$
In addition to the action (38),
it is possible to add terms depending only on a holomorphic
function of $z^i $, the superpotential $g$. These are
$$-{1\over 2}(g_{,ij}\overline{\chi}_i^c \chi^j \ +h.c )
-\vert g_{,i}\vert^2 .  \eqno(41)
$$
The difficulty of generating these terms in the non-commutative
construction is that in order to reproduce the fermionic term
in (41) one must introduce the term $g_{,ij}$ in
the Dirac operator. The bosonic part will then contain terms
of the form $\vert D_{\mu}g_{,ij}\vert^2 =\vert D_{\mu}z^k
g_{,ijk}\vert^2 $ and $\vert g_{,ij}g_,^{*jk}\vert^2 $ which
in general do not coincide with the bosonic part of (38). We
deduce that the terms in an $N=1$ supersymmetric theory
proportional to a general superpotential do not correspond
to a non-commutative action. It would be very interesting
to find out which special forms of the superpotential do correspond
to a non-commutative construction.

To find out whether it is possible to derive the action (38)
from non-commutative geometry we first define a matrix
representation for the spinors $\lambda^a $ and $\chi^i $:
$$
\Psi =\pmatrix{\lambda_{Lj}^i & \chi^i \cr
\chi_j  & 0 \cr \tilde{\lambda_{Li}^j} &\chi_i^T \cr
\chi^{iT} &0\cr}.\eqno(42)
$$
where $\chi_i =C\overline{\chi^i}^T $ is the  right-handed
Weyl spinor conjugate to the left handed $\chi^i $, and
$\tilde{\lambda }_i^j =C\overline{\lambda_i^j}^T $,
$\lambda_j^i =i\lambda^a (T^a)_j^i$  . The reason
we have to take such a complicated representation for the spinors
is due to the fact that we are working with chiral multiplets
which distinguishes between left-handed and right-handed spinors.
The action of the  Dirac operator on the Hilbert space of
these spinors is
$$
D=\pmatrix{\di \ot \delta_i^j & \g5 z_0^i &0 &0\cr
\g5 z_{0i}^* & \di &0 &0 \cr
0&0 &\di \ot \delta_j^i &\g5 z_{0i}^{*T} \cr
0&0&\g5 z_0^{iT} &\di \cr},\eqno(43)
$$
and where the algebra is ${\cal A}=C^{\infty}(M_4)\ot
(M_n(C)\op C )$. The representation $\pi (a)$  is given by
$\pi (a) ={\rm diag }(a, a', \overline{a}, \overline {a'})$
 where $a$ is an $n\times n$ matrix
and $a'$ is a function and the overline denotes complex conjugation.
A one-form $\rho $ has the representation
$$
\pi (\rho )=\pmatrix{A_i^j &\g5 z^{i} &0&0\cr
\g5 z_i^{*} &B &0&0\cr
0&0&\overline{A_i^j} &\g5 z_i^{*T} \cr
0&0&\g5 z^{iT} &\overline {B}\cr},\eqno(44)
$$
where $A_i^j =\sum (a\di b)_i^j $, $B=\sum a'\di b'$,
$z^i+z_0^i =\sum (az_0)^ib'$.
We shall impose the restrictions
$$
A_i^j =iA^a (T^a)_i^j ,\qquad B=0 , \eqno(45)
$$
which reduces the gauge group from $U(N)\times U(1)$
to G.
Next we calculate $\pi (d\rho )$, and this can be found in complete
analogy with eq (21):
$$\eqalign{
\pi (d\rho )_{11}&=\gamma^{\mu\nu}\partial_{\mu} (A_{\nu})_j^i
 +z_j^* z_0^i +z_{0j}^*z^i +\sum \partial^{\mu}a\partial_{\mu}b \cr
\pi (d\rho )_{22}&=\gamma^{\mu\nu}\partial_{\mu}B_{\nu} +
\sum \partial^{\mu}a'\partial_{\mu}b' \cr
\pi (d\rho )_{12}&=\gm \g5 (\partial_{\mu}z^i +(A_{\mu})_j^i z_0^i)
\cr
\pi (d\rho )_{21} &=\pi (d\rho )_{12}^*, \cr}\eqno(46)
$$
and similarly for $\pi (d\rho )_{p,q}$ where $p,q=3,4$ with the
main difference that $z$ and $z^*$ are exchanged. The
other components of $\pi (d\rho )$ vanish.
For $\rho \in {\rm Ker}\pi $, $\pi (d\rho )$  is of the form
$$
\pi (d\rho )={\rm diag}( X_j^i , Y , \overline{X_i^j}^T ) \eqno(47)
$$
Since we have restricted the gauge field $(A_{\mu})_j^i $ to the form (45)
one would expect that a similar constraint must be imposed on
 $X_i^j$. However, it turns out that the only constraint that
would result in the correct bosonic part is
$$
{\rm Tr} (T^a X)=0.
$$
After moding out by the kernel of $\pi (d\rho )$ and redefining
$z+z_0\ra z $, we find that
 $\pi (\theta )$ is:
$$
\pi (\theta )=\pmatrix{{1\over 2}\gamma^{\mu\nu}(F_{\mu\nu})_j^i
+  v_i^j &
\gm\g5 D_{\mu}z^i &0&0\cr
\gm\g5 D_{\mu}z_j^*  & 0 &0&0\cr
0&0& {1\over 2}\gamma^{\mu\nu} (F^*_{\mu\nu})_i^j + v_i^{jT}
 &\gm\g5 D_{\mu}z_i^{*T} \cr 0&0&\gm\g5
D_{\mu}z^{iT} &0 \cr}
\eqno(48)
$$
where
$$ \eqalign{
v_i^j &= ( z^iz_j^* -z_0^iz_0^{j*} )^{\perp}\cr
(z^iz_j^*)^\perp &={1\over \tau_R }(T^a)_j^i (z^*T^az).\cr} \eqno(49)
$$
The choice
of $z_0$ must be such that
$$\eqalign{
(z_0^*T^az_0)&=-\xi^a \qquad ( {\rm for\ abelian\ generators} )\cr
             &=0  \qquad    ({\rm otherwise} )\cr}\eqno(50)
$$
The bosonic part of the non-commutative action becomes:
$$\eqalign{
{1\over 4}{\rm Tr}_w (\theta^2 D^{-4})&=\int \sqrt g d^4x \Bigl(
-{1\over 4}\tau_R F_{\mu\nu}^a F^{\mu\nu a}  \cr
&\qquad +D_{\mu}z_i^* D^{\mu}z^i +{1\over 2}\tau_R \bigl( z_i^* (T^a z)^i
+\xi^a \bigr)^2 \Bigr) \cr}\eqno(51)
$$
while the fermionic action is now simply given by
$$
{1\over 2}\Bigl( \Psi ,[D+\pi (\rho ), \Psi ]\Bigr) \eqno(52)
$$
After the rescaling the full action by $I\ra {1\over \tau_R}I $,
and the fields $z^i $ and $\chi^i $ by $z^i \ra \sqrt {\tau_R} z^i $
 and $\chi^i \ra \sqrt{\tau_R}\chi^i $, one finds out that the
non-commutative action functional which is the sum of (51) and (52),
completely coincides with (38). The supersymmetry transformations,
again take a very simple form
$$\eqalign{
\delta \Psi &= -\pi (\theta )\varepsilon \cr
\delta \pi (\rho )&= (\overline{\varepsilon } E_A \Psi )E_A ,\cr}
\eqno(53)
$$
where
$$
\varepsilon =\pmatrix{\epsilon_L &0 \cr 0&\epsilon_R \cr
\tilde{\epsilon_L} &0 \cr 0&\tilde{\epsilon_R} \cr}, \eqno(54)
$$
The $E_A$ in (53) is an orthonormal basis of  $\Omega_D^1 ({\cal A}) $, which
is now given by
$$\eqalign{
E_a&=\gamma_a \ot \pmatrix{1_n &0\cr 0& 1\cr } \cr
E_i&=i\gamma_5 \ot \pmatrix{0& e_i \cr e_i^* &0 \cr},\cr}\eqno(55)
$$
and $e_i $ is an orthonormal frame in $C^n $ satisfying
$(e_i,e_j)=\delta_j^i $.

It is possible to repeat this excercise for the general coupling
of the $N=2 $ theory, and one finds again that a non-commutative
description is possible. We shall not report on this here, but leave
it for a future publication where a detailed account will be given.

To conclude, we have discovered that extended globally supersymmetric
theories as well as $N=1$ supersymmetric theories without
superpotentials could be derived from the non-commutative
construction. This could be taken as a reinterpretation of the
geometry of supersymmetric theories in the same way that the
standard model admits  such a non-commutative construction.
We have not dealt yet with
the question of whether theories where supersymmetry is spontaneously broken
could be also derived from a non-commutative action functional.
We can immediately say that only very special models may have this
property since we are not allowed to use an arbitrary superpotential.
 Another interesting problem is to study whether
 theories with local supersymmetry (i.e.
including supergravity) could be linked to non-commutative
spaces. All these problems are now under study.

\vskip1truecm
{\bf\noindent Acknowledgments}\hfill\break
I would like to thank J\"urg Fr\"ohlich  for very stimulating
discussions, and Daniel Kastler for his continuing support and
interest.
\vfill
\eject

{\bf \noindent References}
\vskip.2truecm
\item{[1]} A. Connes, {\sl Publ. Math. IHES} {\bf 62} (1983) 44;
{\sl Non-Commutative Geometry }, Academic Press (1993);\br
A. Connes and J. Lott, {\sl in Proceedings of 1991 Cargese
Summer Conference}, eds. J. Fr\"ohlich et al, Plenum Press (1992);
{\sl Nucl. Phys. Proc. Suppl.} {\bf B18} (1989) 29.

\item{[2]} M. Dubois-Violette, R. Kerner, and J. Madore,
{\sl Class. Quant. Grav.} {\bf 6}  (1989) 1709; {\sl J. Math. Phys.}
{\bf 31} (1990) 316;\br
M. Dubois-Violette, {\sl C. R. Acad. Sc. Paris} {\bf 307 I} 403 (1989);\br
J. Madore, {\sl Mod. Phys. Lett.} {\bf A4}  (1989) 2617.
\item{[3]} D. Kastler, {\sl A detailed account of Alain Connes'
version of the standard model in non-commutative geometry} {\bf I,
II, III}, to appear in Rev. Math. Phys.;\br
D. Kastler and M. Mebkhout {\sl Lectures on Non-Commutative
Differential Geometry}, World Scientific, to be published;\br
D. Kastler and T. Sh\"ucker, {\sl Theor. Math. Phys.} {\bf 92}
(1993) 522.

\item{[4]} R. Coquereaux, G. Esposito-Far\'ese and G. Vaillant,
{\sl Nucl. Phys.} {\bf B353} (1991) 689;\br
R. Coquereaux, G. Esposito-Far\'ese and F. Scheck,
{\sl Int. J. Mod. Phys.} {\bf A7} (1992) 6555.

\item{[5]} A.H. Chamseddine, G. Felder and J. Fr\"ohlich,
{\sl Phys. Lett.} {\bf 296B} (1993) 109; {\sl Nucl. Phys.}
{\bf B395} (1993) 672; {\sl Comm. Math. Phys.} {\bf 155} (1993) 205.

\item{[6]} For reviews and references on supersymmetry see:\br
J. Bagger and J. Wess, {\sl Supersymmetry and Supergravity},
Princeton University Press, 1983;\br
P. West, {\sl Introduction to Supersymmetry and Supergravity},
World Scientific, (1986).

\item{[7]} S. Ferrara and B. Zumino, {\sl Nucl. Phys. } {\bf
B79} (1974) 413;\br
A. Salam and J. Strathdee, {\sl Phys. Rev.} {\bf D11} (1975) 1521.

\item{[8]} A. Salam and J. Strathdee, {\sl Phys. Lett.} {\bf
51B} (1974) 353;\br
P. Fayet, {\sl Nucl. Phys.} {\bf B113} (1976) 135.

\item{[9]} L. Brink, J. Schwarz and J. Scherk, {\sl Nucl. Phys. }
{\bf B121} (1977) 77;\br
F. Gliozzi, J. Scherk and D. Olive, {\sl Nucl. Phys.} {\bf B122}
(1977) 253.

\end